\documentclass[conference]{IEEEtran}
\IEEEoverridecommandlockouts
\usepackage[ruled,norelsize]{algorithm2e}
\usepackage{cite}
\usepackage{amsmath,amssymb,amsfonts}
\usepackage{algorithmic}
\usepackage{graphicx}
\usepackage{textcomp}
\usepackage{algorithmic}
\usepackage{graphicx}
\usepackage{xcolor}
\usepackage{amsmath}
\usepackage{amssymb}
\usepackage{bm}
\usepackage{graphicx}
\usepackage{xcolor}
 \usepackage{amsmath}
\usepackage{amsthm}
\usepackage{url}
\usepackage{mathtools}
\usepackage{multirow}

\DeclarePairedDelimiter\floor{\lfloor}{\rfloor}
\theoremstyle{definition}
\newtheorem{definition}{Definition}[section]
\def\BibTeX{{\rm B\kern-.05em{\sc i\kern-.025em b}\kern-.08em
    T\kern-.1667em\lower.7ex\hbox{E}\kern-.125emX}}
    
\makeatletter
\newcommand{\removelatexerror}{\let\@latex@error\@gobble}    
    
\begin{document}

\title{Composition Modulation\\
\thanks{The authors wish to acknowledge the support of the Bristol Innovation $\&$ Research Laboratory of Toshiba Research Europe Ltd. }
}

\author{\IEEEauthorblockN{Ferhat Yarkin and Justin P. Coon}
\IEEEauthorblockA{Department of Engineering Science, University of Oxford, Parks Road, Oxford, UK, OX1 3PJ \\
E-mail: \{ferhat.yarkin and justin.coon\}@eng.ox.ac.uk}
}

\maketitle

\begin{abstract}
In this paper, we propose novel modulation concepts that we call \emph{weak composition modulation} (WCM) and \emph{composition modulation} (CM). We use weak and strict compositions of an integer to form codewords of WCM and CM, respectively. For the proposed schemes, we depict a practical model using orthogonal frequency division multiplexing (OFDM). We also analyze the error performance of the proposed schemes. It is shown that the proposed schemes are capable of outperforming conventional OFDM and OFDM with index modulation (OFDM-IM)  in terms of error performance.
\end{abstract}

\begin{IEEEkeywords}
Integer partitioning, composition, index modulation, orthogonal frequency division multiplexing (OFDM).
\end{IEEEkeywords}

\section{Introduction}

Due to the remarkable growth in mobile data services, the smart devices as well as connectivity, the need for spectrally and energy-efficient, as well as reliable communication systems, increases day-by-day \cite{Gatherer2018}. In this regard, index modulation (IM) has attracted remarkable attention due to its attractive advantages including better error performance and improved energy/spectral efficiency over conventional modulation and multiplexing schemes. Considering the conspicuous success of orthogonal frequency division multiplexing (OFDM) in cellular networks, local area networks, audio broadcasting, and power line networks, the adaptation of IM to OFDM has also attracted several researchers’ attentions \cite{Basar2013,Mao2017,Wen2017, Yarkin2019, Yarkin2020} and it has been shown by these studies that OFDM with IM (OFDM-IM) is capable of achieving better error performance, higher data rate and energy efficiency than OFDM. 

On the other hand, the upcoming 5G networks will support three main services, including enhanced mobile broadband, ultra-reliable low-latency communication, and massive machine type communication, by using different key technologies such as massive multiple-input multiple-output (MIMO), mmWave communications, ultra-dense network and so on \cite{Akyildiz2016}. 5G's key enabling technologies are expected to satisfy the user demands for considerably high data rate, ultra-reliable transmission, and very low latency. However, it is anticipated that 6G and the following next-generation networks will require even more data rate, better error performance and lower latency than what 5G offers due to the proliferation of a variety of new applications, including extended reality services, telemedicine, haptics, flying vehicles, brain-computer interfaces, and connected autonomous systems \cite{Saad2019,Dang2020}. In this regard, the existing OFDM and OFDM-IM schemes require further improvement in error performance and spectral efficiency (SE) to be deployed efficiently in next-generation networks.

Motivated by the advantages of the OFDM-IM schemes as well as the requirements of next-generation networks, we uncover a novel application of a combinatorial tool called composition (or integer composition) in this paper. In this regard, we propose two different modulation concepts, \emph{weak composition modulation} (WCM) and \emph{composition modulation} (CM), that exploit weak and strict compositions of an integer, respectively, to form their codewords. Then, we depict practical OFDM implementations of the proposed schemes and use the compositions to embed information into the energies of the OFDM subcarriers. We also investigate the bit error rate (BER) of the proposed schemes in this paper and obtain an upper bound on the BER. Our analytical, as well as numerical, findings indicate that the proposed schemes can achieve a substantially better
performance than OFDM-IM
and conventional OFDM in terms of BER.

\section{Composition Modulation}

In this section, we describe the basic idea of composition modulation (CM). We begin with some useful definitions and relations. Then, we propose two different modulation concepts based on CM.

\theoremstyle{definition}
\begin{definition}{\textbf{Multiset:}} 
A multiset can be defined as a modified concept of  a set, in which each element is allowed to be repeated a number of times. 
\end{definition}

\theoremstyle{definition}
\begin{definition}{\textbf{Multiset number:}} 
It is the number of multisets of cardinality $I$ that an $N$-set has. To select a multiset of cardinality $I$, we have to select non-negative numbers $\mu_1, \mu_2, \ldots, \mu_N$ such that $\mu_1+\mu_2+\ldots+\mu_N=I$. It can be shown that the number of solutions to that equation is $\binom{I+N-1}{I}$. 
\end{definition}

\theoremstyle{definition}
\begin{definition}{\textbf{Partition:}} 
A partition of a positive integer is a multiset of positive integers that sum to that integer. Hence, a partition of an integer $I$ can be written as
\begin{align}
    I=\mu_1+\mu_2+\ldots+\mu_N
\end{align}
where $\mu_1 \ge \mu_2 \ge \ldots \ge \mu_N \ge 1$ and such a partition is denoted by $(\mu_1, \mu_2, \ldots, \mu_N)$.
\end{definition}

\theoremstyle{definition}
\begin{definition}{\textbf{Composition:}} 
A composition of an integer is an ordered list of positive integers that sum to that integer. Integer composition is sometimes called ordered integer partition. Hence, ordering the elements of an integer partition yields the same partition; however, it yields a different integer composition. It is sometimes called a strict composition since the integers that form a composition are strictly positive. 
\end{definition}

Let $C(I)$ be the total number of compositions of an integer $I$ and let $C_N(I)$ be the number of
compositions of $I$ which have exactly $N$ summands.\footnote{For this case, we say that such a composition has $N$ parts.} One can check that $C(I)=2^{I-1}$ and $C_N(I)=\binom{I-1}{N-1}$. 

\theoremstyle{definition}
\begin{definition}{\textbf{Weak composition:}} A weak composition of an integer is an ordered list of non-negative integers that sum to that integer. Unlike an integer composition, we allow the list elements to be zero for a weak composition. If we do not limit the number of the parts, there are infinitely many weak compositions of an integer.    
\end{definition}

\theoremstyle{definition}
\begin{definition}{\textbf{A weak $N$-composition of $I$:}} A weak $N$-composition of an integer $I$ is an ordered list of $N$ non-negative integers that sum to that integer. The number of weak compositions of $I$ into $N$ parts is $\binom{I+N-1}{N-1}$.    
\end{definition}

\subsection{Weak Composition Modulation}
In weak composition modulation (WCM), a codebook of $L_{WCM}$ codewords $\textbf{x}_1, \ldots, \textbf{x}_{L_{WCM}}$ is formed in a way that each codeword has $N$ elements that are determined by the weak  $N$-compositions of an integer $I$. Hence, when we form $N$-tuples related to the codewords of a WCM codebook, we make the elements in an $N$-tuple equal to the square roots of the elements in a weak $N$-composition of an integer $I$. More explicitly, the $i$th element $x_i$, $i=1,\ldots,N$, of an $N$-tuple is chosen as $x_i=\sqrt{\mu_i}$ where $\mu_i\in \big\{0, 1, \ldots, N\big\}$ is the $i$th summand of the composition.   For example, two weak 4-compositions of $I=4$ can be given as $4=1+1+1+1$ and $4=4+0+0+0$. In this case, the corresponding $N$-tuples are given as $(1, 1, 1, 1)$  for the former and $(2,0,0,0)$ for the latter. Notice that we ensure the same total energy for each $N$-tuple since we incorporate compositions of the same integer throughout the codebook. Moreover, the number of codewords in a WCM codebook can be given as the number of weak $N$-compositions of $I$, i.e., $L_{WCM}=\binom{I+N-1}{N-1}$.


\subsection{Composition Modulation}
In composition modulation (CM), unlike WCM that uses the weak compositions of an integer to form its codewords, we use only strict compositions of an integer to form the CM codewords. Hence, we form the $N$-tuples for CM in a way that the elements are equal to the square roots of the elements of the compositions of an integer $I$ that have $N$ parts. More explicitly, the $i$th element $x_i$, $i=1,\ldots,N$, of an $N$-tuple is chosen as $x_i=\sqrt{\nu_i}$ where $\nu_i\in \big\{ 1, \ldots, I-N+1\big\}$ is the $i$th summand of the composition. Note that one needs to pick $I \ge N$ to be able to construct the CM codewords. The number of codewords, $L_{CM}$, in this scheme, can be given as the number of compositions of an integer $I$ with $N$ parts, i.e., $L_{CM}=\binom{I-1}{N-1}$. Like WCM, we ensure the same total energy for the codewords of CM.

\section{Practical Model For OFDM}

In this section, we present two different system models in which we apply the proposed WCM and CM schemes to OFDM transmissions.

For both of the system models, $m$ input bits enter the transmitter and these bits are separated into $B = m/f$ blocks, each having $f$ input bits. Similarly, the total number of subcarriers $N_T$ is also divided
into $B = N_T /N$ blocks, each having $N$ subcarriers. For each
block of input bits, $f$ information bits are modulated by WCM and CM encoders for OFDM with WCM (OFDM-WCM) and OFDM with CM (OFDM-CM), respectively. The resulting modulated symbols are carried by $N$ subcarriers. 

Since each bit and each subcarrier block has the same mapping operation for the proposed schemes, we focus on a single block, the $b$th block (where $b \in 1, 2, . . . , B$), in the following subsections. In the $b$th block, the $f$ information bits are further divided into two parts, one of them having $f_1$ bits and the other one having $f_2$ bits with $f_1+f_2 = f$. As it will be explained in the next subsections, the remaining operations will be different for the proposed schemes.

\subsection{OFDM with WCM}

In OFDM-WCM, the first $f_1$ bits are used to determine a specific weak $N$-composition of a positive integer $I$ in the $b$th block which is denoted by $\omega^b\coloneqq\big\{\mu_1, \mu_2, \ldots, \mu_N\big\}$ where $\mu_i\in \big\{0, 1, \ldots, I\big\}$. Therefore, $f_1=\floor{\log_2 \binom{I+N-1}{N-1}}$. Then, we map the $i$th element in the selected weak composition to the energy of the $i$th subcarrier in a way that the total energy is preserved. Hence, the energy of the $i$th subcarrier can be given as $E_i=\mu_iE_T/I$ where $E_T=\sum_{i=1}^{N}|x_i^b|^2$ is the total energy of an OFDM block and $x_i^b$ is the symbol carried by the $i$th subcarrier in the $b$th block. Let us denote the set that contains the subcarriers' energies as $\mathcal{E}^b\coloneqq\big\{\mu_1E_T/I, \mu_2E_T/I, \ldots, \mu_NE_T/I \big\}$. Moreover, the mapping of $f_1$ bits
to the subcarriers' energies can be implemented by using a look-up
table. Then, the remaining $f_2$ bits are used to determine $M$-ary symbols on the activated subcarriers. Note that since we utilize weak compositions in this scheme, some of the subcarriers may carry no energy, or in other words, they may be deactivated, depending on the selected weak composition. To send the same amount of information bits for all weak compositions, we adjust the modulation order on the activated subcarriers according to the weak compositions and choose the size of the constellation on the $i$th subcarrier $2^{\lambda \mu_i}$, where $\lambda$ is a positive integer that we use to alter the number of transmitted bits. In that way, we ensure that $f_2=\lambda I$.

Let us consider an example of how we decide the modulation orders on the activated subcarriers and implement the look-up table between the subcarriers' energies and the first $f_1$ information bits for the proposed scheme when $M$-PSK modulation is considered and $N=I=3$ as well as $\lambda=1$. We list the weak 3-compositions of $I=3$ along with the energy sets, $\mathcal{E}_b$, as well as OFDM-WCM codewords in Table \ref{tab:table1}. As seen from the table, we map the weak compositions to the subcarriers and use them to determine the modulation order on the activated subcarriers. 

\begin{table}[t!]
\centering
\caption{Codeword generation and look-up table implementation example for OFDM-WCM when $N=I=3$ and $\lambda=1$.}
\label{tab:table1}
\begin{tabular}{|c|l|l|c|}
\hline
\begin{tabular}[c]{@{}c@{}}Weak \\Compositions\end{tabular} & \multicolumn{1}{c|}{$\mathcal{E}_b$} & \multicolumn{1}{c|}{\begin{tabular}[c]{@{}c@{}}OFDM-WCM\\ Codeword\end{tabular}} & $f_1$ bits \\ \hline
3=0+0+3 & $\big\{0, ~~~~0, ~~~E_T\big\}$ & (0,   0,   8-PSK) & [0 0 0] \\ \hline
3=0+1+2 & $\big\{0, ~~\frac{E_T}{3}, \frac{2E_T}{3}\big\}$ & (0, BPSK, QPSK) & [0 0 1] \\ \hline
3=0+2+1 & $\big\{0, ~\frac{2E_T}{3},~ \frac{E_T}{3}\big\}$ & (0, QPSK, BPSK) & [0 1 0] \\ \hline
3=0+3+0 & $\big\{0,~ ~~E_T,~~~~ 0\big\}$ & (0, 8-PSK, 0) & [0 1 1] \\ \hline
3=1+0+2 & $\big\{\frac{E_T}{3}, ~0, ~\frac{2E_T}{3}\big\}$ & (BPSK, 0, QPSK) & [1 0 0] \\ \hline
3=1+1+1 & $\big\{\frac{E_T}{3}, \frac{E_T}{3}, \frac{E_T}{3}\big\}$ & (BPSK, BPSK, BPSK) & [1 0 1] \\ \hline
3=1+2+0 & $\big\{\frac{E_T}{3}, \frac{2E_T}{3},~~ 0\big\}$ & (BPSK, QPSK, 0) & [1 1 0] \\ \hline
3=2+0+1 & $\big\{\frac{2E_T}{3},~ 0,~ \frac{E_T}{3}\big\}$ & (QPSK, 0, BPSK) & [1 1 1] \\ \hline
3=2+1+0 & $\big\{\frac{2E_T}{3}, \frac{E_T}{3}, ~~0\big\}$ & (QPSK, BPSK, 0) & unused \\ \hline
3=3+0+0 & $\big\{E_T, ~~~~0, ~~~0\big\}$ & (8-PSK, 0, 0) & unused \\ \hline
\end{tabular}
\end{table}

One can map $f_1$ bits to the subcarriers' energies without using a look-up table implementation. More explicitly, a weak composition of an integer can be represented by using dashes and vertical bars similar to a MacMahon graph of a composition \cite{sills2013compositions}. To represent weak $N$-compositions of $I$, $I$ dashes are used to signify the positive parts of the integer, whereas $N-1$ bars are used to indicate the addition operations, or in other words, the separations between the parts. To obtain the dash and bar representation of a composition, we write the dashes first. Then, we place the bars in a way that the $i$th space separated by these bars have $\mu_i$ dashes where $\mu_i$ is the $i$th part of the composition. For example, let's have a look at the weak 3-compositions of $I=3$. In that case, one needs $I=3$ dashes, i.e., $-~ - ~-$, and $N-1=2$ bars since $I=3$ can be composed of three positive parts at most and there will be $N=3$ parts separated by two bars. Hence, the weak compositions $3=1+1+1$ and $3=0+0+3$ can be represented as ``$-|-|-$'' and ``$||---$'', respectively. Hence, each weak composition can be represented by a combination of the bars and dashes. After obtaining the dash and bar representation of a weak composition, the locations\footnote{Note that there are $I+N-1$ locations in the dash and bar representations due to $I$ dashes and $N-1$ bars. For ``$-|-|-$'' and ``$||---$'',  the locations of the bars can be written as $\big\{2, 4\big\}$ and $\big\{1, 2\big\}$, respectively.} of bars or dashes in the representation can be mapped to decimal numbers by using the combinatorial method in \cite{Basar2013}. Finally, the decimal numbers can be converted into binary bits.

The SE of the proposed scheme is given by
\begin{align}
    \eta=\frac{f_1+f_2}{N}=\frac{\floor{\log_2 \binom{I+N-1}{N-1}}+\lambda I}{N}.
\end{align}
By assuming $\mathcal{M}_i$ is a $2^{\lambda\mu_i}$-ary constellation, the OFDM-WCM symbol vector corresponding to the $b$th block can be written as $\textbf{x}^b=[x_1^b, x_2^b, \ldots, x_N^b]$ where $x_i^b\in\mathcal{M}_i$ when $\mu_i \neq 0$, otherwise $x_i^b=0$.  After obtaining symbol vectors for all blocks, an OFDM block creator forms the overall symbol vector  $\textbf{x}\coloneqq [x(1), x(2), \ldots,x(N_T)]^T=[\textbf{x}^1,\ldots, \textbf{x}^b, \ldots, \textbf{x}^B]^T\in \mathcal{C}^{N_T\times 1}$. Note that the energy of the symbol carried by the $i$th subcarrier of the $b$th block can be written as $E_i^b=|x_i^b|^2=\mu_iE_T/I$. After this point, exactly the same operations as conventional OFDM are applied\footnote{We assume that the elements of $\textbf{x}$ are interleaved sufficiently and the maximum spacing is achieved for the subcarriers.}.

At the receiver, the received signal is down-converted, and the cyclic prefix is then removed from each received baseband symbol vector before processing with an FFT. After employing a $N_T$-point FFT operation, the frequency-domain received signal vector can be written as 
\begin{align}\label{eq:eq2}
    \textbf{y} \coloneqq [y(1),y(2), \ldots, y(N_T)]^T=\textbf{X}\textbf{h}+\textbf{n}
\end{align}
where  $\textbf{X}=\text{diag}(\textbf{x})$. Moreover, $\textbf{h}$ and $\textbf{n}$ are $N_T\times 1$ channel and noise vectors, respectively. Elements of these vectors follow the complex-valued Gaussian distributions $\mathcal{CN}(0,1)$ and $\mathcal{CN}(0,N_0)$, respectively, where $N_0$ is the noise variance.

Since the encoding procedure for each block is independent of others, decoding can be performed independently at the receiver. Hence, using maximum likelihood (ML) detection, the detected symbol vector for the $b$th block can be written as 
\begin{align}\label{eq:eq3}
    ({\hat{\mathcal{E}}^b,\hat{\textbf{x}}^b})= \arg \min_{\mathcal{E}^b, \textbf{x}^b} ||\textbf{y}^b-\textbf{X}^b\textbf{h}^b||^2
\end{align}
where $\textbf{y}^b=[y((b-1)N+1), \ldots, y(bN)]^T$, $\textbf{X}^b=\text{diag}(\textbf{x}^b)$ and $\textbf{h}^b=[h((b-1)N+1), \ldots, h(bN)]^T$.

\subsection{OFDM with CM}

In OFDM-CM, the first $f_1$ bits are used to determine a specific composition of a positive integer $I$ with $N$ parts in the $b$th block.  Therefore, $f_1=\floor{\log_2\binom{I-1}{N-1}}$. We map the $i$th element in the selected composition to the energy of the $i$th subcarrier, i.e., $E_i=\nu_iE_T/I$ where $\nu_i \in \big\{1, 2, \ldots, I-N+1\big\}$. Unlike OFDM-WCM, all parts of a composition are positive, thus all subcarriers are activated in OFDM-CM. We denote the set that contains the subcarriers' energies as $\xi^b \coloneqq \big\{\nu_1E_T/I, \nu_2E_T/I, \ldots, \nu_NE_T/I\big\}$. Like OFDM-WCM, the mapping of $f_1$ bits to the set, $\xi^b$, can be done by using a look-up table\footnote{In OFDM-CM, as in OFDM-WCM, one can map $f_1$ bits to the subcarriers' energies without using a look-up table. It is straightforward to show that each strict composition of $I+N$ into $N$ parts has an equivalent weak composition of $I$ into $N$ parts. Such an equivalent weak composition can be obtained by subtracting one from each part of the strict composition.  Hence, after converting the strict composition into a weak composition, the mapping can be done by following the same steps as in the previous subsection.}. Once we decide $\xi^b$ according to the incoming $f_1$ bits, $f_2=N\log_2M$ bits are used to determine $M$-ary constellation symbols carried by the subcarriers, thus the SE of the OFDM-CM scheme can be written as 
\begin{align}
    \eta=\frac{f_1+f_2}{N}=\frac{\floor{\log_2\binom{I-1}{N-1}}+N\log_2M}{N}.
\end{align}
Here, different from OFDM-WCM,
we use the same modulation order, $M$, on each subcarrier since we activate all of the subcarriers during a transmission interval. For brevity, we use the same notation as the OFDM-WCM scheme to denote the OFDM-CM symbol vector. Hence, such a vector for the $b$th block of the OFDM-CM scheme can be written as $\textbf{x}^b=[x_1^b, x_2^b, \ldots, x_N^b]$ where $x_i^b \in \mathcal{M}$ and $\mathcal{M}$ is an $M$-ary constellation. After obtaining symbol vectors of all blocks, the overall OFDM-CM vector is formed as $\textbf{x}\coloneqq [x(1), x(2), \ldots,x(N_T)]^T=[\textbf{x}^1,\ldots, \textbf{x}^b, \ldots, \textbf{x}^B]^T\in \mathcal{C}^{N_T\times 1}$. Note that the energy of the symbol carried by the $i$th subcarrier of the $b$th block can be given as $E_i^b=|x_i^b|^2=\nu_iE_T/I$. Note also that the proposed scheme is equivalent to conventional OFDM when $I=N$. For this case, we have only $\binom{N-1}{N-1}=1$ composition, and $\nu_i=1$, $\forall i \in \big\{1, 2, \ldots, N\big\}$, thus all the symbols carried by the subcarriers have the same energy as in conventional OFDM. After this point, the same transmitter operations as OFDM-WCM are employed. 

At the receiver, by applying the same operations as OFDM-WCM, the received signal can be obtained by substituting the OFDM-CM vector into \eqref{eq:eq2} instead of the OFDM-WCM vector. Moreover, the ML detection can be performed in the same way as \eqref{eq:eq3} and the detected symbol can be written as 
\begin{align}
    ({\hat{\xi}^b,\hat{\textbf{x}}^b})= \arg \min_{\xi^b, \textbf{x}^b} ||\textbf{y}^b-\textbf{X}^b\textbf{h}^b||^2.
\end{align}

\section{Codebook Selection}
It may not always possible to find proper system parameters for the proposed schemes to achieve the same SE as those of OFDM and OFDM-IM schemes. Hence, to provide fair comparisons between the proposed and benchmark schemes, we provide a codebook selection algorithm in this section.  

The optimum codebook selection criterion in a fading channel at high SNR is the well-known rank-determinant criterion \cite{Tarokh1998}. Hence, in the proposed schemes, the error performance at high SNR is limited by the codewords whose difference matrix has the minimum rank. Note that the difference matrix can be defined as $(\textbf{X}^i-\textbf{X}^j)$ for the codeword pairs $\textbf{x}_i$ and $\textbf{x}_j$ where $\textbf{X}^i=\operatorname{diag}(\textbf{x}_i)$ and $\textbf{X}^j=\operatorname{diag}(\textbf{x}_j)$. For the proposed schemes, such a minimum distance is observed from unit rank codeword pairs related to the conventional modulation symbols. To alleviate the effect of these pairs in the codebook,  we propose a practical algorithm that culls these pairs from the codebook until a target SE, $R$, is reached. The proposed algorithm is given in Algorithm \ref{alg:alg1}. The algorithm starts by forming the codebook $\mathcal{C}\coloneqq\big\{\textbf{x}_1, \textbf{x}_2, \ldots, \textbf{x}_L\big\}$ according to one of the proposed schemes where $\textbf{x}_i$ is the $i$th, $i \in \big\{1, 2, \ldots, L\big\}$, codeword in the codebook and $L$ is the size of the codebook. In the second step, the algorithm forms $L \times L$ matrix $\textbf{Z}$ whose $i$th row and $j$th column element, $z_{ij}$, is the rank of the difference matrix $(\textbf{X}^i-\textbf{X}^j)$, thus the function ``rank(.)'' is used to measure the rank of the difference matrices. Then, in the third step, the algorithm culls the codewords that have the maximum number of minimum nonzero rank pairs one-by-one until a target SE, $R$, is reached, i.e., $L=2^R$. In this step, the function ``nonzeros(.)'' is used to find nonzero elements of the matrix $\textbf{Z}$ and assign these elements to a vector, $\textbf{v}$, whereas the function   ``find(.)'' is used to find row indices of elements in the $i$th column of $\textbf{Z}$ that equal the minimum value of $\textbf{v}$, i.e., $z_{min}$. Moreover, $|\textbf{r}_i|$ stands for the cardinality of the vector $\textbf{r}_i$. Then, we determine the codeword that has the maximum number of minimum rank pairs by $\breve{i}=\arg \max_i r_i$ and remove it from the codeook. Finally, we update the codebook size by removing one and the matrix $\textbf{Z}$ by eliminating the $\breve{i}$th row and column of $\textbf{Z}$, i.e., $\textbf{Z}(:,\breve{i})=[~], \textbf{Z}(\breve{i},:)=[~]$.     

\begin{figure}[t]
 \removelatexerror
  \begin{algorithm}[H]
   \caption{Exclusion of minimum rank codeword pairs}
   \label{alg:alg1}
   {\bf Step 1:} Form the codebook $\mathcal{C}$ where $\mathcal{C}\coloneqq\big\{\textbf{x}_1, \textbf{x}_2, \ldots, \textbf{x}_L \big\}$ \;
  {\bf Step 2:} Form $L \times L$ matrix $\textbf{Z}$ where $z_{ij}=\operatorname{rank}(\textbf{X}^i-\textbf{X}^j )$, $i,j \in \big\{1, 2, \ldots, L\big\}$\; 
  {\bf Step 3:} Excluding minimum rank pairs\;
    \While{$L\neq 2^R, \forall \textbf{x}_i\in \mathcal{C}$}
    { 
    $\textbf{v}=\operatorname{nonzeros}(\textbf{Z})$, $z_{min}=\min \textbf{v}$\;
     $\textbf{r}_i=\operatorname{find}(\textbf{Z}(:,i)==z_{min})$, $r_i=|\textbf{r}_i|$\;
    $\breve{i}=\arg \max_i r_i$\;
    $\mathcal{C}\leftarrow \mathcal{C}-\big\{\textbf{x}_{\breve{i}}\big\}$\;
    $L \leftarrow L-1$\;
    $\textbf{Z}(:,\breve{i})=[~]$, $\textbf{Z}(\breve{i},:)=[~]$ \;

}

        {\bf return} $\mathcal{C}$\;
  \end{algorithm}
\end{figure}




\section{Performance Analysis}

An upper-bound on the average BER is given by the well-known union bound as follows
\begin{align}\label{eq:eq9}
    P_b \leq \frac{1}{f2^f}\sum_{i=1}^{2^f}\sum_{j=1}^{2^f}P(\textbf{X}^i\to\textbf{X}^j)D(\textbf{X}^i\to\textbf{X}^j)
\end{align}
where $P(\textbf{X}^i\to\textbf{X}^j)$ is the pairwise error probability (PEP) regarding the erroneous detection of $\textbf{X}^i$ as $\textbf{X}^j$ where $i \neq j$, $i,j\in \big\{1, \ldots,L\big\}$, $\textbf{X}^i=\text{diag}(\textbf{x}^i)$, $\textbf{X}^j=\text{diag}(\textbf{x}^j)$, and $D(\textbf{X}^i\to\textbf{X}^j)$ is the number of bits in error for the corresponding pairwise error event. Here, $L$ is  the codebook size for the proposed schemes. One can use the same PEP expression as in \cite{Basar2013} and substitute the codewords of the proposed schemes to obtain the upper bound on the average BER.

For the proposed schemes, the minimum Hamming distance between the sets that keep the subcarriers' energies is two just like the index symbols of the OFDM-IM schemes. However, the proposed schemes send conventional modulation symbols together with embedding information into the energy domain, and the minimum Hamming distance between the conventional modulation symbols is limited to one. That limits the diversity gain of the proposed schemes to one.       

\section{Numerical Results}\label{sec:secV}

In this section, we provide numerical BER results for the proposed schemes. In figures, ``OFDM-WCM $(I, N, \lambda)$, Alg. \ref{alg:alg1}'' stands for the proposed OFDM-WCM scheme that employs Algorithm \ref{alg:alg1} and uses $\lambda$ to alter the SE as well as integer compositions of $I$ with $N$ parts to decide the energies of $N$ subcarriers, whereas ``OFDM-CM $(I, N, M)$'' stands for the proposed OFDM-CM scheme that determines the energies of $N$ subcarriers according to the composition of an integer $I$ with $N$ parts and carries $M$-ary PSK symbols on each subcarrier. Moreover, ``OFDM-IM $(N, K, M)$'' stands for the conventional OFDM-IM scheme in which $K$ out of $N$ subcarriers are activated to send $M$-PSK modulated symbols in each block.     

\begin{figure}[t!]
		\centering
		\includegraphics[width=8cm,height=6cm]{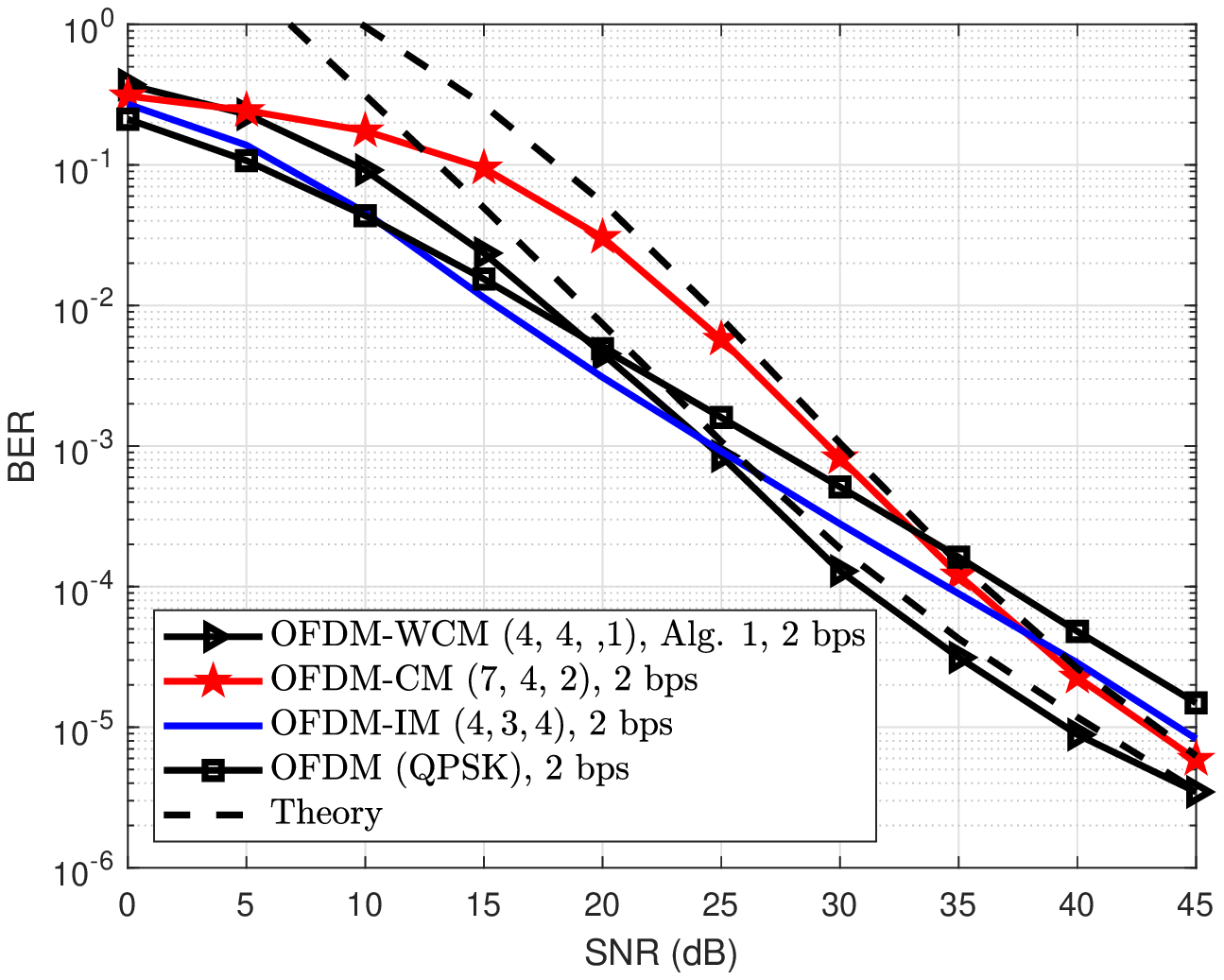}
		\caption{BER comparison of OFDM-WCM $(4, 4, 1)$, Alg. \ref{alg:alg1} and OFDM-CM $(7, 4, 2)$ with OFDM-IM $(4, 3, 4)$ and OFDM (QPSK).}
		\label{fig:fig1}
\end{figure}

 
Fig. \ref{fig:fig1} compares the BER performance of OFDM-WCM $(4, 4, 1)$, Alg. \ref{alg:alg1} and OFDM-CM $(7, 4, 2)$ with OFDM-IM $(4, 3, 4)$ and OFDM (QPSK). Here, the SEs of all schemes are 2 bits per subcarrier (bps). The proposed OFDM-WCM $(4, 4, 1)$ scheme is capable of achieving $\frac{\floor{\log_2\binom{I+N-1}{N-1}}+\lambda I}{N}=\frac{\floor{\log_2\binom{7}{3}}+ 4}{4}=2.25$ bps; however, we cull half of the OFDM-WCM $(4, 4, 1)$ codewords with Algorithm \ref{alg:alg1} to achieve the same SE as those of the remaining schemes. Hence, the remaining codebook, OFDM-WCM $(4, 4, 1)$, Alg. \ref{alg:alg1}, exhibits an SE that is equal to 2 bps. As seen from the figure, OFDM-CM $(7, 4, 2)$ is capable of achieving a better BER performance than OFDM (QPSK) by providing almost the same BER results as those of OFDM-IM $(4, 3, 4)$  at high SNRs. The proposed OFDM-WCM $(4, 4, 1)$, Alg. \ref{alg:alg1} scheme, on the other hand, considerably outperforms all other schemes, especially at high SNRs. In Fig. \ref{fig:fig1}, we also compare the theoretical upper bound results obtained by \eqref{eq:eq9} with the computer simulation results for the BER of the proposed schemes. As seen from the figure, the theoretical results match with the simulation results, especially at high SNRs.

\begin{figure}[t!]
		\centering
		\includegraphics[width=8cm,height=6cm]{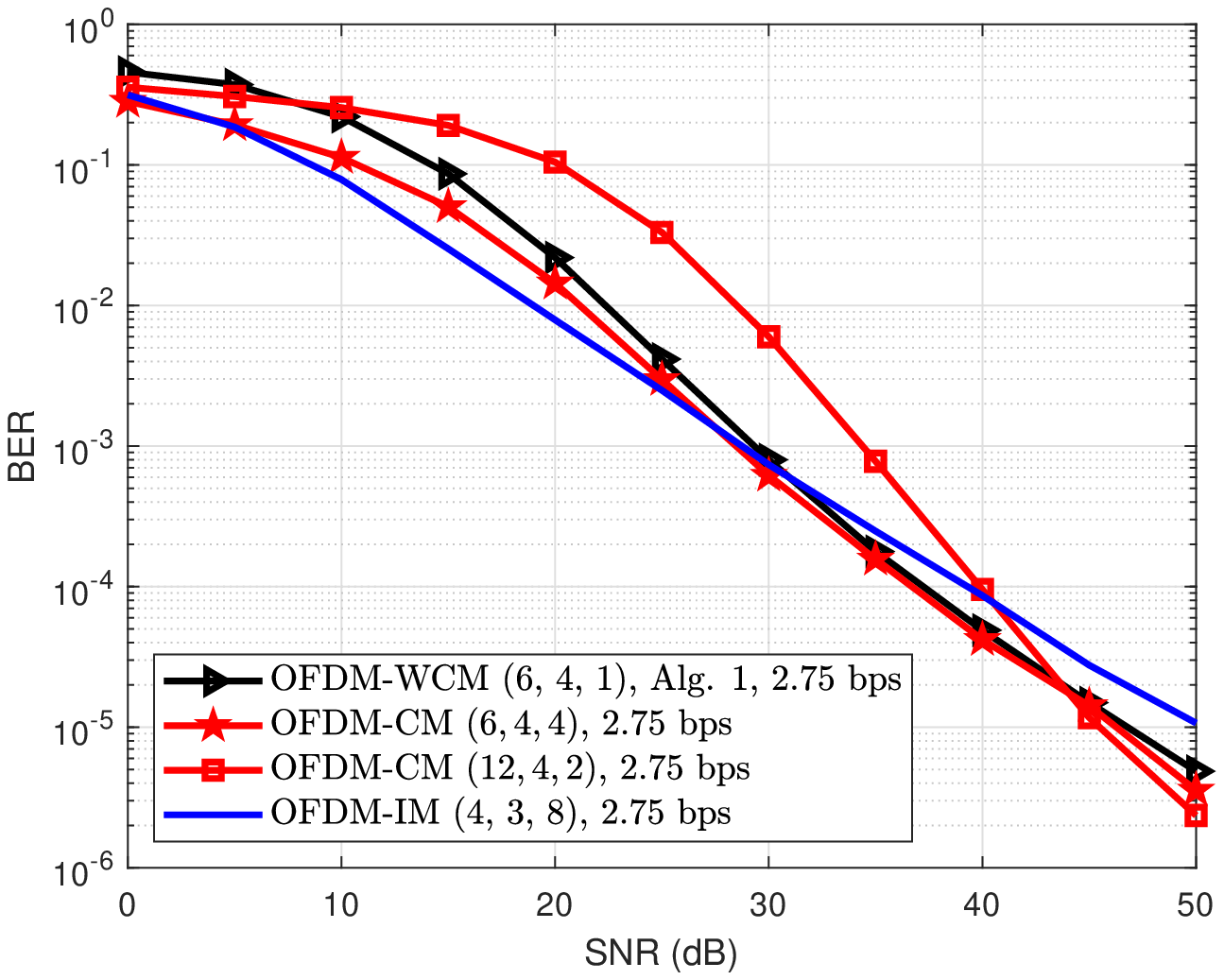}
		\caption{BER comparison of OFDM-WCM $(6, 4, 1)$, Alg. 1, OFDM-CM $(6, 4, 4)$, and OFDM-CM $(12, 4, 2)$ with OFDM-IM $(4, 3, 8)$.}
		\label{fig:fig2}
\end{figure}

In Fig. \ref{fig:fig2}, we compare the BER performance of OFDM-WCM $(6, 4, 1)$, Alg. 1, OFDM-CM $(6, 4, 4)$, and OFDM-CM $(12, 4, 2)$ with OFDM-IM $(4, 3, 8)$. The SEs of all schemes are 2.75 bps\footnote{Here, a block of OFDM-WCM $(6, 4, 1)$ is capable of transmitting $\floor{\log_2\binom{6+4-1}{4-1}}+6=12$ bits; however, we cull the half of the OFDM-WCM $(6, 4, 1)$ codebook by using Algorithm \ref{alg:alg1} to achieve the same SE as a block of OFDM-IM $(4, 3, 8)$ that is capable of sending $11$ bits.}. As seen from the figure, the proposed schemes are capable of outperforming the OFDM-IM scheme especially at high SNRs.  Moreover,  OFDM-WCM $(6, 4, 1)$, Alg. 1, OFDM-CM $(6, 4, 4)$, and OFDM-CM $(12, 4, 2)$ provide almost 3.2 dB, 3.7 dB and 4.5 dB SNR gains, respectively, compared to OFDM-IM $(4, 3, 8)$ at a BER value of $10^{-5}$. 

\section{Conclusion}
In this paper, we proposed novel modulation concepts, which we call WCM and CM. For the proposed modulation concepts, we represented practical implementations in the context of OFDM. We showed through computer simulations and theoretical calculations that the proposed schemes can provide noteworthy improvements compared to conventional OFDM, and OFDM-IM in terms of BER.               

As future work, a low-complexity detector could be designed for the proposed schemes and the proposed CM concept could be concatenated with the OFDM-IM schemes to enhance the SE of these schemes. 

\bibliographystyle{IEEEtran}
\bibliography{invited_paper}

\end{document}